\documentclass{elsart}

\usepackage{natbib}

\usepackage{amssymb}

\begin{document}

\begin{frontmatter}

\title{Comparison of decision tree methods\\ for finding active
objects}

\author{Yongheng Zhao \and Yanxia Zhang}

\address{National Astronomical
Observatories, CAS, 20A Datun Road, Chaoyang District, Bejing
100012 China}

\begin{abstract}
The automated classification of objects from large catalogues or
survey projects is an important task in many astronomical surveys.
Faced with various classification algorithms, astronomers should
select the method according to their requirements. Here we describe
several kinds of decision trees for finding active objects by
multi-wavelength data, such as REPTree, Random Tree, Decision Stump,
Random Forest, J48, NBTree, AdTree. All decision tree approaches
investigated are in the WEKA package. The classification performance
of the methods is presented. In the process of classification by
decision tree methods, the classification rules are easily obtained,
moreover these rules are clear and easy to understand for
astronomers. As a result, astronomers are inclined to prefer and
apply them, thus know which attributes are important to discriminate
celestial objects. The experimental results show that when various
decision trees are applied in discriminating active objects
(quasars, BL Lac objects and active galaxies) from non-active
objects (stars and galaxies), ADTree is the best only in terms of
accuracy, Decision Stump is the best only considering speed, J48 is
the optimal choice considering both accuracy and speed.

\end{abstract}


\end{frontmatter}

\section{Introduction}
\label{sect:intro} 

With the development and deployment of a variety of large surveys,
including 2MASS (the Two Micron All Sky Survey), SDSS (the Sloan
Digital Sky Survey), DENIS (the Deep Near Infrared Survey), DIVA ,
GAIA, etc., astronomical data are measured by Terabyes, even
Petabytes. How to automatedly collect, save, process, analyze the
huge amount of database is an important task for astronomers. To
meet the need, different methods are developed. For example, Neural
networks (NN) has been employed for spectral classification of stars
(Storrie-Lombardi et al. 1994; Gulati et al. 1994), for physical
measurement of star spectra (Bailer-Jones et al. 1997), for spectral
classification of galaxies (Sodr$\acute{e }$\& Cuevas 1994), for
morphological classification of galaxies (Storrie-Lombardi et al.
1992; Adams \& Woolley 1994) and for discriminating stars and
galaxies in digitized photographic plates (Odewahn \& Nielsen 1994),
for fast cosmological parameter estimation (Auld et al. 2007) and
for separating quasars from stars (Zhang \& Zhao 2007). Support
vector machines (SVMs) have also been successfully applied to
automatic classification (Zhang \& Zhao 2003, 2004), object
detection (Qu et al. 2003), identification of red variables
(Williams et al. 2004) and redshift estimation (Wadadekar 2005).
Decision trees were applied for building an online system for
automated classification of X-ray sources (McGlynn et al. 2004), for
star-galaxy classification problems (Djorgovski et al. 1994; Fayyad
et al. 1993) and for star-galaxy Classification (Ball et al. 2006).

Each technique has its pros and cons. The largest merit of NN
methods is that they are general: they can deal with problems with
high dimensions and even with complex distributions of objects in
the $n$-dimensional parameter space. NN is becoming popular in
astronomy due to its associated memory characteristic and
generalization capability. However, the relative importance of
potential input variables, long training process, and interpretative
difficulties have often been criticized. Although SVM has high
performance in classification problems (Zhang \& Zhao, 2003), the
rules obtained by SVM algorithm are hard to understand directly.
Moreover, one possible drawback of SVM is its computational cost.

Owing to the above-mentioned drawbacks of NN and SVM, the purpose of
this study is to explore the performance of classification using
various decision tree approaches. Decision tree methods exhibit the
capability of modeling complex relationship between variables
without strong model assumptions. Besides, unlike NN, they are able
to identify ``important" independent variables through the built
tree and basis functions when many potential variables are
considered. Thirdly, decision trees do not need a long training
process and hence can save lots of modeling time when the data set
is huge. Finally, one strong advantage of decision trees over other
classification techniques is that the resulting classification model
can be easily interpreted. They not only point out which variables
are important in classifying objects/observations, but also indicate
that a particular object/observation belongs to a specific class
when the built rules are satisfied. The final fact has important
implications and can help astronomers make better decisions. To be
more clear, the advantages of decision tree methods are listed as
follows:

(1) Decision trees are easy to understand;

(2) Decision trees are easily converted to a set of production
rules;

(3) Decision trees can classify both categorical and numerical data,
but the output attribute must be categorical;

(4) There are no a priori assumptions about the nature of the data.

Certainly, decision tree algorithms also have their disadvantages.
For instance, multiple output attributes are not allowed. They are
unstable. Slight variations in the training data can cause different
attribute selections at each choice point within the tree. The
effect can be significant since attribute choices affect all
descendent subtrees. Trees created from numeric data sets can be
quite complex since attribute splits for numeric data are binary. It
is the rules for splitting that population at the nodes that are
simple, but that there can be large numbers of nodes if the tree is
not pruned. However, when researchers want to obtain clear rules or
distinguish which parameters influence the classification results,
they are inclined to choose decision trees.

Since decision trees have the described advantages, they have proven
to be effective tools in handling forecasting and classification
problems (McGlynn, et al. 2004; Zhang \& Zhao, 2007). In this paper
we describe and apply some decision tree methods for separating
active objects from multiband data. Section 2 introduces decision
tree methods. The sample is indicated in Section 3. The experiment
and discussion are given in Section 4. Finally Section 5 summarizes
the results.

\section{Methods}

Decision trees represent a supervised approach to classification. A
decision tree is a simple structure where non-terminal nodes
represent tests on one or more attributes and terminal nodes reflect
decision outcomes. The ordinary tree consists of one root, branches,
nodes (places where branches are divided) and leaves. In the same
way the decision tree consists of nodes which stand for circles, the
branches stand for segments connecting the nodes. A decision tree is
usually drawn from left to right or beginning from the root
downwards, so it is easier to draw it. The first node is a root. The
end of the chain ``root - branch - node-...- node" is called
``leaf". From each internal node (i.e. not a leaf) may grow out two
or more branches. Each node corresponds with a certain
characteristic and the branches correspond with a range of values.
These ranges of values must give a partition of the set of values of
the given characteristic.

Decision trees we study are from WEKA (The Waikato Environment for
Knowledge Analysis). WEKA is a tool for data analysis and includes
implementations of data pre-processing, classification, regression,
clustering, association rules, and visualization by different
algorithms. A book describing the software was published in 2005 by
Ian H. Witten and Eibe Frank (Witten and Frank, 2005). WEKA's
binaries and sources are freely available. Implemented methods
include instance-based learning algorithms, statistical learning
like Bayes methods and tree-like algorithms like ID3 and J4.8
(slightly modified C4.5). Including combinations of classifiers,
e.g. bagging and boosting schemes, there are over sixty methods
available in WEKA.

The following gives the short introduction of various decision tree
algorithms.

\subsection{REPTree}

REPTree is a fast decision tree learner which builds a
decision/regression tree using information gain as the splitting
criterion, and prunes it using reduced-error pruning. It only sorts
values for numeric attributes once. Missing values are dealt with
using C4.5's method of using fractional instances.

\subsection{RandomTree}

With $k$ random features at each node., a random tree is a tree
drawn at random from a set of possible trees. In this context ``at
random" means that each tree in the set of trees has an equal chance
of being sampled. Another way of saying this is that the
distribution of trees is ``uniform".  Random trees can be generated
efficiently and the combination of large sets of random trees
generally leads to accurate models. Random tree models have been
extensively developed in the field of Machine Learning in the recent
years.

\subsection{J48}

J48 is slightly modified C4.5 in WEKA. The C4.5 algorithm generates
a classification-decision tree for the given data-set by recursive
partitioning of data. The decision is grown using Depth-first
strategy. The algorithm considers all the possible tests that can
split the data set and selects a test that gives the best
information gain. For each discrete attribute, one test with
outcomes as many as the number of distinct values of the attribute
is considered. For each continuous attribute, binary tests involving
every distinct values of the attribute are considered. In order to
gather the entropy gain of all these binary tests efficiently, the
training data set belonging to the node in consideration is sorted
for the values of the continuous attribute and the entropy gains of
the binary cut based on each distinct values are calculated in one
scan of the sorted data. This process is repeated for each
continuous attributes. For a deeper introduction of this method,
readers can refer to (Mitchell 1997; Quinlan 1986).

\subsection{DecisionStump}

A decision stump is basically a one-level decision tree where the
split at the root level is based on a specific attribute/value
pair.

\subsection{Random Forest}

Random forest (Breiman, 2001) is an ensemble of unpruned
classification or regression trees, induced from bootstrap samples
of the training data, using random feature selection in the tree
induction process. Prediction is made by aggregating (majority vote
for classification or averaging for regression) the predictions of
the ensemble. Random forest generally exhibits a substantial
performance improvement over the single tree classifier such as CART
and C4.5. It yields generalization error rate that compares
favorably to Adaboost, yet is more robust to noise.

\subsection{NBTree}

The naive Bayesian tree learner, NBTree (Kohavi 1996), combined
naive Bayesian classification and decision tree learning. In an
NBTree, a local naive Bayes is deployed on each leaf of a
traditional decision tree, and an instance is classified using the
local naive Bayes on the leaf into which it falls. The algorithm for
learning an NBTree is similar to C4.5. After a tree is grown, a
naive Bayes is constructed for each leaf using the data associated
with that leaf. An NBTree classifies an example by sorting it to a
leaf and applying the naive Bayes in that leaf to assign a class
label to it. NBTree frequently achieves higher accuracy than either
a naive Bayesian classifier or a decision tree learner.

\subsection{ADTree}

The alternating decision tree (ADTree) is a generalization of
decision trees, voted decision trees and voted decision stumps. A
general alternating tree defines a classification rule as follows.
An instance defines a set of paths in the alternating tree. As in
standard decision trees, when a path reaches a decision node it
continues with the child which corresponds to the outcome of the
decision associated with the node. However, when reaching a
prediction node, the path continues with all of the children of
the node. More precisely, the path splits into a set of paths,
each of which corresponds to one of the children of the prediction
node. We call the union of all the paths reached in this way for a
given instance the ``multi-path" associated with that instance.
The sign of the sum of all the prediction nodes that included in a
multi-path is the classification which the tree associates with
the instance. The basic algorithm can refer to Freund \& Mason
(1999).

\section{Sample}

We adopted the same sample as that in Zhang \& Zhao (2004). The
sample include the multiwavelength data of 3,718 stars, 173 normal
galaxies, 909 quasars, 135 BL Lacs and 612 active galaxies from
optical (USNO A-2.0), X-ray (The ROSAT Bright Source and Faint
Source ) and infrared bands (2MASS). The chosen attributes from
different bands are $B-R$ (optical color), $B+2.5log(CR)$, $logCR$
(source count-rate in the broad energy band), $HR1$ (hardness ratio
1), $HR2$ (hardness ratio 2), $ext$ (source extent), $extl$
(likelihood of source extent), $J-H$ (infrared color), $H-K_{\rm s}$
(infrared color), $J+2.5log(CR)$.  In the following sections, AGNs
represent quasars, BL Lacs and active galaxies, non-AGNs for stars
and normal galaxies.

\section{Experiment and Discussion}

We conduct experiments to compare various decision tree algorithms
which are implemented within the WEKA framework (Witten and Frank,
2005). We use the implementation of REPTree, Random Tree, Decision
Stump, Random Forest, J48, NBTree and AdTree in WEKA with default
parameters. In our experiment, the accuracy on the sample has been
obtained using 10-fold cross validation, which is helpful to prevent
overfitting. In the following, accuracy is an average of any $9/10$
sample as training set and the rest as testing set for 10 times.
Missing values are also processed using the mechanism in WEKA.

For REPTee, the number of trees to create a classifier is 4,305. In
the case of RandomTree, the number adds up to 90,699. DecisionStump
chooses $H-K_{\rm s}$ as the standard attribute for classification.
When $ H-K_{\rm s} \le 0.3285$ and $H-K_{\rm s}$ is missing, objects
are identified as non-AGNs, while $H-K_{\rm s}> 0.3285$, objects are
classified as AGNs. For J48, we used pruning and a confidence factor
of 0.25. In the case of RandomForest, 10 trees were used for
creating the forest for the experiments, each constructed while
considering 4 random features. Out of bag error (Breiman, 2001) is
0.086. For ADTree, tree size (total number of nodes) is 31, leaves
(number of predictor nodes) is 21. All the experiment results are
shown in Table 1 and Table 2. Table 1 shows the number of correct
instances and incorrect instances, the accuracy of AGNs and
Non-AGNs. Table 2 gives the number of trees to build models, the
accuracy and the time to build each model by 10-fold cross
validation for the individual classifiers, respectively.

\begin{table*}[ht]
\begin{center}
\caption{The classification results for various decision trees}
\bigskip
\begin{tabular}{rllll}
\hline
 Methods & No. of correct & No. of incorrect& Accuracy&Accuracy\\
  & instances&instances&of AGNs&of Non-AGNs\\
\hline
 REPTree & 4410 &1137 &33.15\% &99.23\%\\
 RandomTree & 4850 &697 &64.07\%&97.38\%\\
 DecisionStump & 5282 & 265&93.36\% &96.02\%\\
 RandomForest & 5375 & 172 &93.06\%&98.54\%\\
 J48 &5383 &164 &93.60\% & 98.51\%\\
 NBTree &5392 & 155&95.17\% & 98.07\%\\
 ADTree & 5397 & 150& 95.29\% & 98.15\%\\
 \hline
 \end{tabular}
\bigskip
\end{center}
\end{table*}

\begin{table*}[ht]
\begin{center}
\caption{The performance for various decision Trees }
\bigskip
\begin{tabular}{rllll}
\hline
Methods  & No. of Trees & Accuracy &Time to build models (seconds)\\
\hline
 REPTree & 4305 &79.50\%&6.70\\
 RandomTree & 90699 &87.43\%&5.33\\
 DecisionStump & 1 & 95.22\%&0.09\\
 RandomForest & 10 &96.90\%&70.72\\
 J48 &41 &97.04\% &0.53\\
 NBTree &37 & 97.21\% &60.42\\
 ADTree & 31 & 97.30\% &1.53\\
 \hline
 \end{tabular}
\bigskip
\end{center}
\end{table*}

From Table 1, REPTree and RandomTree show better performance to
classify Non-AGNs, but poor performance in separating AGNs. For the
number of Non-AGNs is more than two times as that of AGNs, REPTree
and RandomTree are easy to obtain rules from large datasets, so both
the two methods are not fit to deal with imbalanced samples. As
shown by Table 2, the rank of accuracy for these decision trees is
ADTree (97.30\%), NBTree (97.21\%), J48 (97.04\%), RandomForest
(96.90\%), DecisionStump (95.22\%), RandomTree (87.43\%), REPTree
(79.50\%). The performance of ADTree, NBTree, J48 and RandomForest
is comparable. REPTree is the most inferior. Of all the decision
tree methods, DecisionStump has highest speed in building the model
and takes 0.09 s while RandomForest is the slowest model requiring
70.72 s for the same. Considering both accuracy and speed, ADTree
and J48 are the best choices.

The accuracy obtained by decision tree method as presented here is
somewhat inferior to the earlier reported accuracy (Zhang \& Zhao
2004) of 97.80\% by Learning Vector Quantisation (LVQ), 98.05\% by
Single Layer Perceptron (SLP), and 98.31\% by Support Vector
Machines (SVMs). But the rules obtained by decision tree algorithms
are clear and easy to understand, so astronomers are inclined to
employ them and know which attributes are important, thus may choose
good features to describe the physics of celestial objects.

\section{Conclusion}

We briefly reviewed and implemented decision tree methods (i.e.
REPTree, Random Tree, Decision Stump, Random Forest, J48, NBTree and
AdTree) in the WEKA framework, focusing on the problem of
differentiating AGN candidates from non-AGN. Decision Stump, Random
Forest, J48, NBTree and AdTree show better performance for our
problem (more than 95.00\%), but REPTree and Random Tree are also
useful and may be better fit to deal with other problems. In our
case, ADTree shows the best performance only in terms of accuracy,
Decision Stump is the best only considering speed, J48 is the
optimal choice considering both accuracy and speed. In the process
of knowledge discovery, choice of parameters and the construction of
high quality training/test data sets are important steps. The large
survey projects are in urgent need of automated classification
systems. Apparently several methods can not meet the requirements of
astronomical research due to the quantities, quality and complexity
of astronomical data. Therefore various techniques are required to
test and employ in order to get reliable classifications. Not only
supervised methods should be tried, but also unsupervised methods
and other methods especially for outlier detection should be
experimented. In addition, the ensembles of some methods are needed.
For example, when facing difficulty in applying neural network
algorithms in high dimensional spaces, some data preprocessing may
be considered. On these occasions, feature selection/extraction
methods may be used for reducing dimensions or noise. Future work
includes testing these methods for other types of astronomical
objects, such as nebulas and clusters, or for other types of data,
for instance, images and spectra. These methods can be used for
redshift measurement, physical parameter measurement of celestial
objects, or morphology
classification of galaxies, and also for feature selection.\\

{\bf Acknowledgments} This paper is funded by National Natural
Science Foundation of China under grant No.10473013 and
No.90412016.


\begin{thebibliography}{}

\bibitem[Adams(94)]{label 1}
Adams, A., Woolley, A., Hubble classification of galaxies using
neural networks. Vistas in Astronomy 38, 273-280, 1994.
\bibitem[Auld(07)]{label 2}
Auld, T., Bridges, M., Hobson, M.P., Fast cosmological parameter
estimation using neural networks, MNRAS 376, L11-L15, 2007.
\bibitem[Ball(06)]{label 3}
Ball, N.M., Brunner, R.J., et al., Robust Machine Learning Applied
to Astronomical Data Sets. I. Star-Galaxy Classification of the
Sloan Digital Sky Survey DR3 Using Decision Trees, ApJ 650, 497-509,
2006.
\bibitem[Breiman(01)]{label 4}
Breiman, L., Random forests, Machine Learning 45(1), 5-32, 2001.
\bibitem[Djorgovski(94)]{label 5}
Djorgovski, Weir, N., and Fayyad, U., Processing and analysis of the
Palomar-STScI Digital Sky Survey using a novel software technology.
In Crabtree, D. R. Hanisch, R. J., and Barnes, J., editors,
Astronomical Data Analysis Software and Systems III, pp.195-204, San
Francisco, ASP, 1994.
\bibitem[Fayyad(93)]{label 6}
Fayyad, U.M., Weir, N., and Djorgovski, S., SKICAT: A machine
learning system for automated cataloging of large scale sky surveys.
In Proceeding of the Tenth International Conference on Machine
Learning, Amherst, MA, Morgan Kaufman, 112-119, 1993.
\bibitem[Freund(99)]{label 7}
Freund, Y., Mason, L., The alternating decision tree learning
algorithm. Proceeding of the Sixteenth International Conference on
Machine Learning, Bled, Slovenia, 124-133, 1999.
\bibitem[Odewahn(94)]{label 8}
Odewahn, S.C., Nielsen, M. L., Star-galaxy separation using neural
networks. Vistas in Astronomy 38, 281-285, 1994.
\bibitem[Kohavi(96)]{label 9}
Kohavi, R., Scaling up the cccuracy of naive-Bayes classifiers: a
decision-tree hybrid, Proceedings of the Second International
Conference on Knowledge Discovery and Data Mining (KDD-96), AAAI
Press 202-207, 1996.
\bibitem[Mitchell(97)]{label 10}
Mitchell, T. Machine Learning. McGraw Hill, 1997.
\bibitem[Quinlan(86)]{label 11}
Quinlan, J.R. Induction of decision trees. Machine Learning 1(1)
81-106, 1986.
\bibitem[Storrie(94)]{label 12}
Storrie-Lombardi, M.C., Irwin, M.J., von Hippel, T., et al.:
Spectral classification with principal component analysis and
artificial neural networks. Vistas in Astronomy 38(3) 331-340, 1994.
\bibitem[Gulati(94)]{label 13}
Gulati, R.K., Gupta, R., Gothoskar, P., et al.: Ultraviolet stellar
spectral classification using multilevel tree neural network, Vistas
in Astronomy 38, 293-299, 1994.
\bibitem[Bailer-Jones(97)]{label 14}
Bailer-Jones, C.A.L., Irwin, M., von Hippel, T., Physical
parametrization of stellar spectra: the neural network approach,
MNRAS 292, 157-166, 1997.
\bibitem[Sodr(94)]{label 15}
Sodr$\acute{e}$, L.Jr., Cuevas, H., Spectral classification of
galaxies, Vistas in Astronomy 38, 286-291, 1994.
\bibitem[McGlynn(04)]{label 16}
McGlynn, T.A., Suchkov, A.A., Winter, E.L., et al., Automated
classification of ROSAT sources using heterogeneous multiwavelength
source catalogs, ApJ 616-1284, 2004.
\bibitem[Storrie(92)]{label 17}
Storrie-Lombardi, M.C., Lahav, O., Sodr, L., et al., Morphological
classification of galaxies by artificial neural networks. MNRAS 259,
8-12, 1992.
\bibitem[Wadadekar(05)]{label 18}
Wadadekar, Y., Estimating photometric redshifts using support vector
machines, PASP 117(827), 79-85, 2005.
\bibitem[Williams(04)]{label 19}
Williams, S.J., Wo$\acute{z}$niak, P. R., Vestrand, W. T., et al.
Identifying Red Variables in the Northern Sky Variability Survey, AJ
128, 2965-2976, 2004.
\bibitem[Witten(99)]{label 20}
Witten, I.H., and Frank E., Data Mining: Practical Machine Learning
Tools and Techniques with Java Implementations, Morgan Kaufmann, San
Francisco, 2005.
\bibitem[Qu(03)]{label 21}
Qu, M., Shih, Frank Y., Jing, J., et al., Automatic solar flare
detection using MLP, RBF, and SVM, Solar Physics 217(1), 157-172,
2003.
\bibitem[Zhang(03)]{label 22}
Zhang, Y., Zhao, Y., Classification in Multidimensional Parameter
Space: Methods and Examples, PASP 115, 1006-1018, 2003.
\bibitem[Zhang(04)]{label 23}
Zhang, Y., Zhao, Y., Automated clustering algorithms for
classification of astronomical objects, A\&A 422, 1113-1121, 2004.
\bibitem[Zhang(07)]{label 24}
Zhang, Y., Zhao, Y., A Comparison of BBN, ADTree and MLP in
separating Quasars from Large Survey Catalogues, ChJAA 7, 289-296,
2007.
\end{thebibliography}
\end{document}